\documentclass[letter, twocoulmn]{IEEEtran} 

\usepackage[dvips]{graphicx}
\usepackage{amsmath,amssymb,amsfonts}
\usepackage[para]{threeparttable}
\usepackage{color}

\usepackage{amssymb}
\usepackage{footnote}
\usepackage{tablefootnote}

\usepackage{algorithm}
\usepackage{algcompatible}
\usepackage{algpseudocode}
\usepackage{cite}
\usepackage{array}
\usepackage{eqparbox}

\newtheorem{proposition}{Proposition}

\newtheorem{remark}{Remark}

\begin{document}

\title{A Desired PAR-Achieving Precoder Design for Multi-User MIMO OFDM based on \\Concentration of Measure 
}
%
%
%
\author{Hyun-Su~Cha,~\IEEEmembership{Student Member,~IEEE,} and  
         Dong Ku~Kim,~\IEEEmembership{Senior Member,~IEEE,}
     
\thanks{H.-S. Cha and D. K. Kim (\textit{Corresponding author}) are with the school of Electrical and Electronic Engineering, Yonsei University, Seoul 120-749, Korea. (e-mail: hyunsu.cha0@gmail.com, dkkim@yonsei.ac.kr). 

}

}
\maketitle  


%

\begin{abstract}
For multi-user multiple-input and multiple-output (MIMO) wireless communications in orthogonal frequency division multiplexing systems, we propose a MIMO precoding scheme providing a desired peak-to-average power ratio (PAR) at the minimum cost that is defined as received SNR degradation. By taking advantage of the concentration of measure \cite{Michel:Concentration,Donoho:HighDimension}, we formulate a convex problem with constraint on the desired PAR. Consequently, the proposed scheme almost exactly achieves the desired PAR on average, and asymptotically attains the desired PAR at the $10^{-3}$ point of its complementary cumulative distribution function, as the number of subcarriers increases.

\end{abstract}


\section{Introduction}

\IEEEPARstart{M}{ultiple-input multiple-output} (MIMO) and orthogonal frequency division multiplexing (OFDM) have been regarded as key technologies for wireless communication systems to boost the network capacity. Regretfully, the OFDM has a fundamental drawback of high peak-to-average power ratio (PAR) \cite{Larsson:M_MIMO_Megazine} that can be further aggravated by MIMO precoding techniques \cite{PAPR_Intensified_By_MIMO}, and it may ultimately result in the use of linear power amplifiers in spite of their high cost \cite{Aggarwal:jd}.


Under this practical challenge, the authors of \cite{Studer:en} and \cite{Cha:eb} effectively used redundant spatial dimensions to considerably reduce the PAR, while providing the inherent MIMO precoding gain for the multi-user (MU) multiple-input single-output and the single-user MIMO, respectively.

In this paper, we propose a MU-MIMO precoding scheme for OFDM system that can achieve a desired PAR performance at the minimum cost by utilizing the redundant spatial dimensions. The cost is defined as the received SNR degradations, compared to those of SNRs that would have been obtained by the original block diagonalization (BD) scheme \cite{BD:Heath_Jr,CB_Chae:BD_VecPerturbation}.

For this purpose, we modify the BD precoding matrix with the introduction of design parameters for the beamforming and cost control, and formulate their relation as a quadratic over linear expression motivated by an ellipsoid constraint in \cite{Cha:eb}. By taking advantage of the concentration of measure \cite{Michel:Concentration,Donoho:HighDimension}, meaning that most of the volume of a high dimensional convex body are concentrated near its boundary, we clarify that the average power consumption hardly changes with respect to the design parameters. Then it offers an opportunity for an accurate approximation of the PAR measure, which is reflected in a convex constraint to guarantee the desired PAR.





The volume concentration phenomenon is intensified as the number of subcarriers $(K)$ increases, which makes the approximated PAR more accurate. As a consequence, the proposed scheme almost exactly achieves the desired PAR on average, and asymptotically attains the desired PAR at the $10^{-3}$ point of its complementary cumulative distribution function (CCDF), as $K$ increases.\footnote{Throughout this paper, $\mathbb N,\mathbb R,\mathbb R_+$ and $\mathbb C$ denote the sets of natural numbers, real numbers, positive real numbers and complex numbers, respectively. For $n\in\mathbb{N}$, denote $[1:n]=\{1,...,n-1,n\}$. For a set of matrices $\{{\mathbf{A}}_k\}$, ${\sf blkdiag}(\mathbf{A}_1,...,\mathbf{A}_n)$ and ${\cal N}(\mathbf{A}_k)$ respectively denote the block diagonal matrix composed of $\{\mathbf{A}_k\}$ and a set of basis vectors on the null space of $\mathbf{A}_k$ except for zero vector.}

\section{System Model and Preliminaries}\label{sec:system_model}
\begin{figure}[!t]
  \centering
   \includegraphics[width=3.10 in]{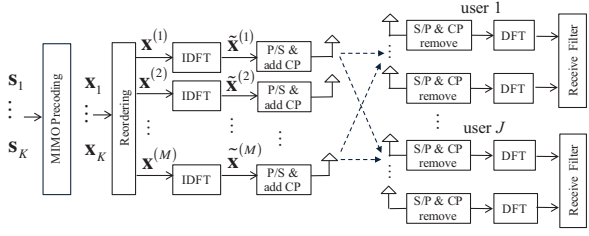}
   \caption{Multi-user MIMO OFDM system.}
   \label{Fig1}
 \end{figure}

In this section, we state the system model and the related studies. Consider the MU-MIMO OFDM system depicted in Fig. \ref{Fig1}, where a transmitter and each user are equipped with $M$ and $N$ antennas respectively, where $(J-1)N<M$ for $J\ge2$ and $N\le M$ for $J=1$. The transmitter wishes to send a signal vector ${\mathbf{s}}_{k}^{[j]}\in\mathbb C^{d_j}$ to the $j^{th}$ user through the $k^{th}$ subcarrier for $j\in[1:J]$ and $k\in[1:K]$, where ${\mathbf{s}}_{k}^{[j]}$ consists of $d_j$ data streams drawn from a discrete constellation set ${\cal A}$, and $d_\Sigma=\sum\nolimits_{j = 1}^J {{d_j}}\le M$.


Let ${\mathbf{H}}_k^{[j]}\in\mathbb{C}^{N\times M}$, ${\bf{F}}_k^{[j]}\in {\mathbb C^{{M} \times d_j}}$, ${\bf{R}}_k^{[j]}\in{\mathbb C}^{d_j\times N}$ and ${\mathbf{w}}_k^{[j]} \in {\mathbb{C}^{{N}}}$ respectively denote the channel matrix composed of i.i.d. complex coefficients, the precoding matrix, the receiving filter and the additive noise that follows circularly symmetric complex Gaussian ${\cal{C}}{\cal{N}}(\mathbf{0},\sigma^2\mathbf{I})$ for all $j\in[1:J]$ and $k\in[1:K]$. We assume that the transmitter completely knows $\{{\mathbf{H}}_k^{[j]}\}_{k\in[1:K],j\in[1:J]}$. Then the received signal vector of the $j^{th}$ user, denoted by ${\mathbf{y}}_k^{[j]}\in\mathbb C^{d_j}$, is given by 
\begin{equation}\label{Input_Output}
{\mathbf{y}}_k^{[j]} = \sum\nolimits_{l = 1}^J { {\mathbf{R}}_k^{[j]} {\mathbf{H}}_k^{[j]}{\mathbf{F}}_k^{[l]}{\mathbf{s}}_k^{[l]}} + {\mathbf{R}}_k^{[j]}{\mathbf{w}}_k^{[j]}
\end{equation}
for all $k\in[1:K]$ and $j \in [1:J]$. 

Let us denote ${{\mathbf{s}}_k} = {[ {{\mathbf{s}}_k^{[ 1],T},\ldots,{\mathbf{s}}_k^{[J],T}}]}^T\in\mathbb{C}^{d_\Sigma\times1}$, ${{\mathbf{F}}_k} = [ {{\mathbf{F}}_k^{[1]}, \ldots ,{\mathbf{F}}_k^{[ J]}} ]\in\mathbb C^{M\times d_\Sigma}$ and ${\bf{x}}_k = \mathbf{F}_k\mathbf{s}_k$ where ${\sf E}[{{{\mathbf{s}}_k}{\mathbf{s}}_k^H} ] = \frac{{{{\cal P}_s}}}{{{d_\Sigma }}}{{\mathbf{I}}_{{d_\Sigma }}}$ for all $k\in[1:K]$, and define $\mathbf{s}= {[ {{\mathbf{s}}_1^T, \ldots ,{\mathbf{s}}_{{K}}^T}]}^T\in\mathbb C^{Kd_\Sigma}$, ${\mathbf{F}} ={\sf blkdiag}({{{\mathbf{F}}_1}, \ldots ,{{\mathbf{F}}_{{K}}}})\in\mathbb C^{MK\times Kd_\Sigma}$ and $\mathbf{x}={[ {{\mathbf{x}}_1^T, \ldots ,{\mathbf{x}}_{{K}}^T} ]}^T\in\mathbb C^{MK\times1}$, where ${\sf E}[{{{\mathbf{s}}}{\mathbf{s}}^H} ] = \frac{{{{\cal P}_s}}}{{{d_\Sigma }}}{{\mathbf{I}}_{{Kd_\Sigma }}}$ and ${\sf E}[ {\left\| {{\mathbf{x}}} \right\|^2} ] \leq K{{\cal{P}}_s}$. 
 All elements in $\mathbf{x}$ are distributed to the $M$ transmission antennas through an one-to-one mapping matrix $\mathbf{\Phi}\in\mathbb R^{MK\times MK}$ in the reordering block of Fig. \ref{Fig1}. This process is also given by \cite{Studer:en} and is represented as
\begin{equation}\label{ordering}
{[ {{{\mathbf{x}}^{( 1),T}}, \ldots ,{{\mathbf{x}}^{(M),T}}}]}^T=\mathbf{\Phi}{[ {{\mathbf{x}}_1^T, \ldots ,{\mathbf{x}}_{{K}}^T} ]}^T,
\end{equation}
where $\mathbf{x}^{(i)}$ is the signal vector transmitted by the $i^{th}$ antenna, and $\mathbf{\Phi}$ is one of the permutation matrices, where their rows are composed of standard basis vectors of $\mathbb R^{1\times MK}$.

Let ${{{\tilde{\mathbf{x}}}}}^{(i)}={{\mathbf{Q}}_i}{{\mathbf{x}}}^{(i)}$ be time-domain signals transmitted by the $i^{th}$ antenna, where ${{\mathbf{Q}}_i}$ is the $K$-point inverse discrete Fourier transform (IDFT) matrix. We define ${\mathbf{Q}}={\sf blkdiag}({{\mathbf{Q}}_1},\ldots,{{\mathbf{Q}}_{M}})$, and assume that the channel tap length is always shorter than the cyclic prefix length. 

Then the PAR of the $i^{th}$ transmission antenna is defined as 
\begin{align}\label{eqs:per_ant_par}
\text{PAR}^{(i)} = {{K\| {{{\tilde {\mathbf{x}}}^{(i)}}}\|_\infty ^2} \mathord{\left/
 {\vphantom {{K\| {{{\tilde{\mathbf{ x}}}^{(i)}}} \|_\infty ^2} {{{\| {{{\tilde{\mathbf{ x}}}^{(i)}}}\|}^2}}}} \right.
 \kern-\nulldelimiterspace} {{{\| {{{\tilde{\mathbf{x}}}^{(i)}}} \|^2}}}} 
\end{align}
for all $i \in [1:M]$. Even if $\mathbf{\Phi}$ can be regarded as a design parameter, we consider a simplified model by assuming $\mathbf{\Phi}=\mathbf{I}$ similar to \cite{Studer:en}. Let us define a relaxed PAR measure in terms of a single linear system with regard to $\mathbf{x}\in\mathbb C^{MK}$ as
\begin{align}\label{PAR_Linear}
{\text{PAR}_{ L}} = \frac{ \|\tilde{\mathbf{ x}}\|_\infty^2}{{\|\tilde{\mathbf{x}}\|^2}\frac{1}{MK}}=\frac{ \|\mathbf{Q x}\|_\infty^2}{{\|\mathbf{Q x}\|^2}\frac{1}{MK}},
\end{align}
where $\tilde{\mathbf{x}}={\mathbf{Qx}}$, and \eqref{PAR_Linear} is also given by \cite{Studer:en,Cha:eb}.

As a preliminary to this paper, we briefly introduce the BD precoding scheme in the following remark \cite{BD:Heath_Jr,CB_Chae:BD_VecPerturbation}.


\begin{remark}[original BD scheme]\label{Block_diagonalization}
For $j\in[1:J]$ and $k\in[1:K]$ where $J\ge2$, let us define $\overline {\mathbf{H}} _k^{[j]} \in {\mathbb{C}^{\left( {J - 1} \right)N \times M}}$ as
\begin{align}
\overline {\mathbf{H}} _k^{[ j]} = {\left[{\mathbf{H}}_k^{[ 1],T}, \ldots ,{\mathbf{H}}_k^{[ {j - 1}],T},{\mathbf{H}}_k^{[ {j + 1}],T}, \ldots, {\mathbf{H}}_k^{[ J],T}\right]^T}, 
\end{align}
$q=M-JN+N$ and ${\bf{U}}_k^{[j]}\in\mathbb C^{M\times q}$ satisfying ${\overline {\mathbf{H}}} _k^{[j]}{\bf{U}}_k^{[j]}=\mathbf{0}$. By singular value decomposition, let us represent ${\mathbf{H}}_k^{[j]}{\bf{U}}_k^{[j]} = {\mathbf{L}}_{k,L}^{[j]}[{\mathbf{\Lambda}}_k^{[j]},\mathbf{0}_{N\times(q-N)}]{\mathbf{L}}_{k,R}^{[j],H}$ where ${\mathbf{\Lambda}}_k^{[j]}={\sf diag}(\lambda_{k,1}^{[j]},\ldots,\lambda_{k,N}^{[j]})$, and $\{\lambda_{k,l}^{[j]}\}_{l=1}^{N}$ denote singular values of ${\mathbf{H}}_k^{[j]}{\bf{U}}_k^{[j]}$. ${\mathbf{V}}_k^{[j]}\in\mathbb C^{q\times d_j}$ is the first $d_j$ columns of ${\mathbf{L}}_{k,R}^{[j]}$, and ${{\bf{R}}_k^{[j]}}$ is the first $d_j$ rows of ${\mathbf{L}}_{k,L}^{[j],H}$. Then BD precoder is ${\bf{U}}_k^{[j]} {\mathbf{V}}_k^{[j]}$. For $J=1$, assume $\overline{\mathbf{H}}_k^{[j]}=\mathbf{H}_k^{[j]}$ and ${\bf{U}}_k^{[j]} =\mathbf{I}$, and determine the precoding matrix and ${\bf{R}}_k^{[j]}$ in the same way.
\end{remark}

In \cite{Aggarwal:jd}, the $l_\infty$-norm minimization technique is introduced as an alternative way to reduce the PAR due to non-convexity of the PAR measure. It is actively used to develop MIMO precoding schemes, such as \cite{Studer:en} for $N=1$ and $J<M$ and \cite{Cha:eb} for $J=1$ and $N<M$. In this work, our main focus is not the PAR reduction by $l_\infty$-norm minimization, but the achivement of the desired PAR performance at the minimum cost for an arbitrary $J$, $M$ and $N$ where $(J-1)N< M$, and also discuss the impact of a large $M$ and $K$.

\section{MU-MIMO Precoder Design}\label{sec:precoder_design}

In this section, we specifically describe the proposed MIMO precoding scheme and discuss its related issues.







\subsection{Modified BD and Design Parameter Definition}\label{subsec:Modified_BD}


We first design the basic structure of the proposed precoding matrix in a bottom-up approach. Let $\gamma\in(0,1]$ denote the received SNR reduction ratio compared to the original BD scheme in Remark \ref{Block_diagonalization}. Then the received SNR for each data stream can be represented as 
\begin{align}
\gamma\cdot{{{\cal P}_s}{{| {\lambda _{k,l}^{[j]}} |^2}}}({d_\Sigma\sigma^2})^{-1}
\end{align}
for all $l\in[1:d_j]$, $j\in[1:J]$ and $k\in[1:K]$, and the cost for achieving the desired PAR is an amount of the SNR reduction compared to Remark \ref{Block_diagonalization}, i.e., $(1-\gamma) {{{\cal P}_s}{{| {\lambda _{k,l}^{[j]}} |^2}}}\frac{1}{{d_\Sigma\sigma^2}}$.
For a given $\gamma$ and $\{d_j\}_{j=1}^{J}$ satisfying $\sum_{j=1}^Jd_j<M$, define
\begin{align}
{{\mathbf{\dot F}}_k}=\sqrt{\gamma}\left[ { {\mathbf{U}}_k^{[ 1]}{\mathbf{V}}_k^{[ 1]}, \ldots ,{\mathbf{U}}_k^{[ J]}{\mathbf{V}}_k^{[ J]}} \right]\in\mathbb{C}^{M\times d_\Sigma}\label{F_dot}
\end{align}
for all $k\in[1:K]$, where ${\mathbf{U}}_k^{[j]}$ and ${\mathbf{V}}_k^{[j]}$ are given by Remark \ref{Block_diagonalization}. Then let us fix the number of redundant spatial dimensions to use to attain the desired PAR by determining
\begin{align}
c_j\in\left[1:M-\left(J-1\right)N-d_j\right],\label{c_j}
\end{align}
and choose $c_j$ vectors in ${\cal N}( {{\mathbf{R}}_k^{[ j]}{\mathbf{H}}_k^{[ j]}{\mathbf{U}}_k^{[ j]}})$, which construct the column vectors of ${\mathbf{P}}_k^{[j]}\in\mathbb C^{q\times c_j}$ for all $j\in[1:J]$ and $k\in[1:K]$. We denote ${c_\Sigma } = \sum\nolimits_{j=1}^Jc_j$, and introduce an arbitrary matrix ${\mathbf{T}}_k=\left[\mathbf{t}_{k,1},\ldots,\mathbf{t}_{k,d_\Sigma}\right]\in\mathbb C^{c_\Sigma\times d_\Sigma}$ as a free design parameter, which is carried by
 \begin{align}
{{\mathbf{\ddot F}}_k}=\left[ {{\mathbf{U}}_k^{[ 1 ]}{\mathbf{P}}_k^{[ 1]}, \ldots ,{\mathbf{U}}_k^{[ J]}{\mathbf{P}}_k^{[ J]}} \right]\in\mathbb{C}^{M\times c_\Sigma} \label{F_double_dot}
\end{align}
for all $k\in[1:K]$. Based on \eqref{F_dot}, \eqref{F_double_dot} and $\mathbf{T}_k$, we construct the proposed precoding matrix structure as
\begin{align}
{{\mathbf{F}}_k} ={{\mathbf{\dot F}}_k}+{{\mathbf{\ddot F}}_k}\mathbf{T}_k
\end{align}
for all $k\in[1:K]$. 

For $K$ subcarriers, let us denote ${{\mathbf{\dot F}}},{{\mathbf{\ddot F}}}$ and $\mathbf{T}$ by block diagonal matrices composed of $\{{{\mathbf{\dot F}}_k}\}_{k=1}^K,\{{{\mathbf{\ddot F}}_k}\}_{k=1}^K$ and $\{\mathbf{T}_k\}_{k=1}^K$ respectively, and construct ${{\mathbf{F}}} ={{\mathbf{\dot F}}}+{{\mathbf{\ddot F}}}\mathbf{T}$.

\begin{remark}
$\{c_j\}_{j\in[1:J]}$ of \eqref{c_j} are related to the redundant spatial dimensions between the transmitter and the $j^{th}$ user, which determine the free variable size with $d_\Sigma$. 
\end{remark}

\begin{remark}
$\{\mathbf{T}_k\}_{k=1}^K$ are used to attain the desired PAR by beamforming on the null space of the effective channels, where ${\mathbf{R}_k^{[l]}}{{\mathbf{H}}} _k^{[l]}{\mathbf{U}}_k^{[j]}{\mathbf{P}}_k^{[j]}=\mathbf{0}$ for all $j,l\in[1:J]$ and $k\in[1:K]$. 
\end{remark}

We derive an equivalent from of $\tilde{\mathbf{x}}=\mathbf{Q}({\dot{\mathbf{{F}}}}+{\ddot{\mathbf{F}}}\mathbf{T})\mathbf{s}$ as a vector expression of $\mathbf{T}$. Let ${{\mathbf{ t}}_k} = {[ {{\mathbf{ t}}_{k,1}^T, \ldots ,{\mathbf{ t}}_{k,d_\Sigma}^T} ]^T}\in{\mathbb{C}^{{c_\Sigma } d_\Sigma\times1}}$ be a vector expression of ${\mathbf{T}}_k$ for all $k\in[1:K]$. Then ${{{\mathbf{\ddot F}}}_k}{{\mathbf{T}}_k}{{\mathbf{s}}_k}$ can be rewritten in terms of $\mathbf{t}_k$, as follows.
 \begin{equation}
\begin{gathered}
     \left[ {{\mathbf{U}}_k^{[ 1]}{\mathbf{P}}_k^{[1]},\ldots,{\mathbf{U}}_k^{[J]}{\mathbf{P}}_k^{[J]}} \right]\left[ {{s_{k,1}}{{\mathbf{I}}_{{c_\Sigma }}},\ldots,{s_{k,d_\Sigma}}{{\mathbf{I}}_{{c_\Sigma }}}} \right] \hfill \\
  {\text{ \,\,   }} \times {\left[ {{\mathbf{t}}_{k,1}^T,\ldots,{\mathbf{t}}_{k,d_\Sigma}^T} \right]^T} = {{\mathbf{G}}_k}{{\mathbf{t}}_k}, \hfill \\ 
\end{gathered} 
\end{equation}       
and ${{\mathbf{x}}_k}= \sum\nolimits_{j = 1}^J { \sqrt{\gamma}\,{\mathbf{U}}_k^{[j]}{\mathbf{V}}_k^{[j]}{\mathbf{s}}_k^{[j]}}+{{\mathbf{G}}_k}{{\mathbf{t}}_k}$ for all $k\in[1:K]$. For $K$ subcarriers, define ${\mathbf{ t}} = {\left[ {{\mathbf{t}}_1^T, \ldots ,{\mathbf{ t}}_{{K}}^T} \right]^T}\in\mathbb C^{c_\Sigma d_\Sigma K\times1}$,
\begin{align}
&\mathbf{G}={\sf blkdiag}\left({\mathbf{G}}_1,\ldots,{\mathbf{G}}_{K}\right)\in\mathbb{C}^{MK\times c_\Sigma d_\Sigma K},\label{G_mat}\\
&{\mathbf{b}} = \left[ {\begin{array}{*{20}{c}}
  {\sum\nolimits_{j = 1}^J {{\mathbf{U}}_1^{[j]}{\mathbf{V}}_1^{[j]}{\mathbf{s}}_1^{[j]}} } \\ 
   \vdots  \\ 
  {\sum\nolimits_{j = 1}^J {{\mathbf{U}}_{{K}}^{[j]}{\mathbf{V}}_{{K}}^{[j]}{\mathbf{s}}_{{K}}^{[j]}} } 
\end{array}} \right]\in\mathbb{C}^{MK\times1}.\label{b_vec}
\end{align}
As a result, we can represent $\tilde{\mathbf{x}}=\mathbf{Q}({{\mathbf{Gt}} + \sqrt{\gamma}\,{\mathbf{b}}})$.



\subsection{Convex Optimization for the Proposed Precoder Design}\label{subsec:par_measure}






Let us formulate the convex problem for the design parameters, $\mathbf{t}$ and $\gamma$. We assume that $\|\mathbf{Fs}\|^2$ does not highly varies depending on $\mathbf{s}$, if $K$ is large enough such that $|{\cal A}|\ll Kd_\Sigma$. Motivated by this, we average out the denominator of \eqref{PAR_Linear} with respect to $\mathbf{s}\in\mathbb C^{kd_\Sigma}$. Then, ${{\sf E}_{\mathbf{s}}[\| \tilde{\mathbf{x}}\| ^2]}  =    \gamma K {\cal P}_s + {\sf{E}}[ {{{ \| {\ddot{\mathbf{F}}\mathbf{Ts}}\|^2}}} ]\le  \gamma K {\cal P}_s +\frac{ {\cal P}_s }{d_\Sigma}{\sf tr}(\mathbf{T}\mathbf{T}^H)=\gamma K {\cal P}_s + \frac{ {\cal P}_s }{d_\Sigma}\|\mathbf{t}\|^2$, and we can define $\gamma K {\cal P}_s +\frac{ {\cal P}_s }{d_\Sigma}\|\mathbf{t}\|^2\le K{\cal P}_s$ from $\gamma\le1$.

From this aspect, we define a PAR measure as a function of $\mathbf{t}$ and $\gamma$ through the relaxation of \eqref{PAR_Linear}, as follows.
\begin{align}\label{Relaxed_Target_PAR_LB}
\overline{\text{PAR}}_{L}=\frac{MK {\| \mathbf{Q}\left({{\mathbf{Gt}} + \sqrt{\gamma}\,{\mathbf{b}}}\right)   \|_\infty^2 } }{  \gamma K {{\cal P}_s}  + \frac{ {\cal P}_s }{d_\Sigma}\|\mathbf{t}\|^2}.
\end{align}
Especially for a fixed $\gamma=\hat\gamma$, the set of feasible $\mathbf{t}$ is defined as a following geodesic ball.\begin{equation}\label{Norm_Ball}
{\cal B}_\mathbb{C}:={\cal B}_\mathbb{C}(r)=\{\mathbf{t}\in\mathbb{C}^{ m } : \left\|\mathbf{t}\right\|^2\le r^2 \},
\end{equation}
where $m={{c_\Sigma } d_\Sigma K}$ and $r^2=d_\Sigma K(1-\hat\gamma)$. 

In \eqref{Norm_Ball}, $1-\hat\gamma$ is reflected in the radius $r$, and the volume of the ball depends on $r$ for a fixed $m$. Notice that the feasible space of $\mathbf{t}$ is described as the inner space enclosed by the ball including the boundary, and its size is the volume of the ball. Thus, the feasible space size is directly affected by $1-\hat\gamma$.

The resultant PAR by the proposed scheme could be fairly well represented by \eqref{Relaxed_Target_PAR_LB}, while it is not a convex function with respect to $\mathbf{t}$ and $\gamma$. We state a proposition by taking advantage of the concentration of measure \cite{Michel:Concentration,Donoho:HighDimension,John_Info_Age}, which gives us the key idea for the convex approximation of \eqref{Relaxed_Target_PAR_LB}.

\begin{proposition}\label{Const_Measure}
For a given $c_\Sigma$, $d_\Sigma$ and $\gamma=\hat\gamma$, 
\begin{align}\label{prop:Approx_PAR}
\overline{\text{PAR}}_{L}\simeq{ {M\|  {\mathbf{Q}}({{\mathbf{Gt}} + \sqrt{\hat\gamma}\,{\mathbf{b}}})  \|_\infty^2} \mathord{\left/
 {\vphantom { {M\|  {\mathbf{Q}}({{\mathbf{Gt}} + \sqrt{\hat\gamma}\cdot{\mathbf{b}}})  \|_\infty^2} {{\cal P}_s}}} \right. \kern-\nulldelimiterspace} {{\cal P}_s}}
\end{align}
is satisfied with respect to most of the feasible $\mathbf{t}$ in $\{\mathbf{t}\in\mathbb R^m : {\| {{\mathbf{t}}} \|^2} \le d_\Sigma K(1-\hat\gamma)\}$, if $K\in\mathbb N$ is arbitrarily large.
\end{proposition}

\begin{IEEEproof}
Let ${\cal B}_\mathbb{R}$ denote the set $\{\mathbf{t}\in\mathbb R^m : {\| {{\mathbf{t}}} \|^2} \le d_\Sigma K(1-\hat\gamma)\}$, and its volume is denoted by ${\sf vol}({\cal B}_\mathbb{R})$.
From \cite[Ch.2]{John_Info_Age}, we define $\epsilon\in\mathbb R_+$ where $\epsilon\ll1$, and denote $(1-\epsilon){\cal B}_{\mathbb R}=\{(1-\epsilon)\mathbf{t}:\mathbf{t}\in {\cal B}_{\mathbb R}\}$, and then the volume ratio is expressed as $\frac{{\sf vol}((1-\epsilon){\cal B}_{\mathbb R})}{ {\sf vol}({\cal B}_{\mathbb R})}=(1-\epsilon)^m$. For a large $m$, ${\sf vol}((1-\epsilon){\cal B}_{\mathbb R})\ll{\sf vol}({\cal B}_{\mathbb R})$ holds, even if $\epsilon$ is arbitrarily small. It means that most of the volume is concentrated near the boundary, and most of the feasible space of $\mathbf{t}$ also exists near the boundary, i.e., most of the feasible $\mathbf{t}$ satisfy $\|\mathbf{t}\|^2\simeq d_\Sigma K(1-\hat\gamma)$ if $K$ is arbitrarily large. Finally, substitute this into \eqref{Relaxed_Target_PAR_LB}, then \eqref{prop:Approx_PAR} is derived.
 \end{IEEEproof}

\begin{remark}\label{remark:complex_ball}
Even if the exact volume comparison between ${\cal B}_\mathbb{C}(r_1)$ and ${\cal B}_\mathbb{C}(r_2)$ is too difficult \cite[eq. (11)]{Barg:Sphere_Exact}, ${\cal B}_\mathbb{C}$ also shows the volume concentration property, which is shown in Talagrand concentration inequality \cite[Theorem 2.1.13]{tao2012_topics} and the statistical study \cite{dryden:2005}. 
\end{remark}


From Remark \ref{remark:complex_ball}, we can infer that Proposition \ref{Const_Measure} is also valid for $\mathbf{t}\in\mathbb{C}^{m}$. Then let us formulate the convex problem.


\textit{1) Objective function}: We minimize $1-\gamma$ for the variable $\gamma$ to minimize the received SNR reduction by the proposed scheme compared to Remark \ref{Block_diagonalization}.





\textit{2) Constraints}: In case $\gamma$ is a variable, the relation between $\mathbf{t}\in\mathbb C^m$ and $\gamma\in\mathbb R_+$ is expressed as a quadratic over linear function, denoted by $\|\mathbf{t}\|^2\leq d_\Sigma K (1-\gamma)$. Let a real function $f:\mathbb C^{m}\to\mathbb R$ denote a mapping by the square of $l_2$-norm, then $g\left( {{\mathbf{t}}\in\mathbb C^m,1-\gamma\in\mathbb R_+} \right)=(1-\gamma) f\left(\frac{\mathbf{t}}{(1-\gamma)}\right)$ is a convex function by \cite{Compex_Conv:Heinz}. For a fixed $\gamma$, the feasible space of $\mathbf{t}$ is not expressed as the quadratic over linear constraint but is represented as a high dimensional $l_2$-norm ball, where most of its volume is concentrated near the surface, and hence $\|\mathbf{t}\|^2\simeq d_\Sigma K(1-\gamma)$ still holds for the overall feasible $\gamma$.

Let us define $\zeta\in\mathbb R_+$ as the desired PAR value, where $\zeta>1$, and represent $\overline{\text{PAR}}_{L} \simeq { {M\|  {\mathbf{Q}}({{\mathbf{Gt}} + \sqrt{\gamma}\,{\mathbf{b}}})  \|_\infty^2} \mathord{\left/ {\vphantom { {M\|  {\mathbf{Q}}({{\mathbf{Gt}} + \sqrt{\gamma}\,{\mathbf{b}}})  \|_\infty^2} {{\cal P}_s}}} \right. \kern-\nulldelimiterspace} {{\cal P}_s}} = \zeta$ from Proposition \ref{Const_Measure}. For convex relaxation, we approximate $\sqrt \gamma \simeq \frac{1}{2}\left( {1 + \gamma } \right)$ assuming $\gamma>0.5$, which can be considered as a quite reasonable approach, since our interest is $\gamma$ of near 1 and this approximation becomes more tight as $\gamma$ goes to 1. Based on these, we state the follwoing convex problem.
\begin{align}
({\cal C})\begin{cases}\mathop {{\text{minimize}}}\limits_{{\mathbf{t}}\in\mathbb C^{m},\gamma\in\mathbb R_+ } &1 - \gamma,\\
\mbox{ subject to }& 0.5<\gamma\le1,\\
&{{{\left\| {\mathbf{t}} \right\|}^2} \le d_\Sigma{K}\left( {1 - \gamma } \right),}\\
&{\left\| {\mathbf{Q}\left( {{\mathbf{Gt}} + 0.5\left( {1 + \gamma } \right) {\mathbf{b}} }\right) } \right\|_\infty } \le \sqrt{\frac{\zeta{\cal P}_s}{M}}, \nonumber
 \end{cases}
\end{align}
where $m=c_\Sigma d_\Sigma K$ and $\zeta>1$.


For all $j\in[1:J]$ and $k\in[1:K]$, the overall design procedure is described as follows. \textbf{Step 1)} compute ${\bf{U}}_k^{[j]},{\bf{V}}_k^{[j]}$ and ${\bf{R}}_k^{[j]}$ from Remark \ref{Block_diagonalization}, and determine $c_j$ and ${\mathbf{P}}_k^{[j]}$ by referring to Section \ref{subsec:Modified_BD}. Construct $\ddot{\mathbf F}_k$ described in \eqref{F_double_dot}. \textbf{Step 2)} find the solution $(\hat\gamma,\hat {\mathbf{t}})$ of $({\cal C})$ by using interior point methods (IPMs) \cite{Boyd:st, Compex_Conv:Heinz}, and then construct $\dot{\mathbf F}_k$ by using $\hat\gamma$, as shown in \eqref{F_dot}. \textbf{Step 3)} decompose $\hat{\mathbf{t}}$ to $\hat{\mathbf{t}}_k={{{[ {{\mathbf{t}}_{k,1}^T, \ldots ,{\mathbf{t}}_{k,{d_\Sigma }}^T}]}^T}}$, and make ${\hat{\mathbf{T}}_k}=[ {\hat{\mathbf{t}}_{k,1}, \ldots ,\hat{\mathbf{t}}_{k,{d_\Sigma }}^{}}]$ from $\hat{\mathbf{t}}_k$. Finally, we construct $\hat{\mathbf{T}}$, $\dot{\mathbf{F}}$ and $\ddot{\mathbf{F}}$ as block diagonal matrices composed of $\{{\hat{\mathbf{T}}_k}\}_{k=1}^K$, $\{{\dot{\mathbf{F}}_k}\}_{k=1}^K$ and $\{{\ddot{\mathbf{F}}_k}\}_{k=1}^K$ respectively, and determine ${\hat{\mathbf{F}}}=\dot{\mathbf{F}}+\ddot{\mathbf{F}}\hat{\mathbf{T}}$.



\subsection{Discussion}\label{subsec:discussion}

We discuss several issues related to the proposed scheme with an assumption of $\gamma=\hat\gamma$.

\textit{1) Effect of a large $K$}: For $\mathbf{t}\in\mathbb R^m$ and $r^2=d_\Sigma K (1-\hat\gamma)$, most of the volume of ${\cal B}_\mathbb{R}(r)$ exists in an annulus of width ${O}\left(\sqrt{\frac{1-\hat\gamma}{{c_\Sigma^2 d_\Sigma K}}}\right)$ near the boundary if $c_\Sigma^2 d_\Sigma K$ is arbitrarily large by \cite[Ch.2]{John_Info_Age}. This indicates that the volume concentration phenomenon is intensified as $K$ increases, and hence the feasible space of $\mathbf{t}$ is also concentrated near the boundary as $K$ increases. As a result, $\|\mathbf{t}\|^2\simeq (1-\hat\gamma)d_\Sigma K$ becomes more solid value with respect to the overall feasible $\mathbf{t}$, and naturally \eqref{prop:Approx_PAR} becomes more accurate.






{\textit{2) User-specific cost and computational complexity}}: Let $\alpha_j\in(0,1]$ denote the received SNR reduction ratio of the $j^{th}$ user. Intuitively, we can determine $\{\alpha_j\}_{j=1}^J$ with constraint on $\hat\gamma=\frac{1}{d_\Sigma}\sum\nolimits_{j=1}^J \alpha_jd_j$ to distribute the total cost to $J$ users by considering the received SNR strength, required quality of service and etc. Secondly, nonlinear convex problems are generally solved by IPMs \cite{Boyd:st, Compex_Conv:Heinz}. One of them, primal-dual infeasible path-following method requires the computational complexity of ${\cal{O}}(m^3)$ where $m=c_\Sigma d_\Sigma K$.






\textit{3) Effect of a large $M$ for a fixed $d_\Sigma$}: More redundant spatial dimensions do not always provide the better PAR performance, as stated in the following proposition.

\begin{proposition}\label{Massive_MIMO}
For ${\cal B}_\mathbb{C}$ in (\ref{Norm_Ball}), if $c_\Sigma\to\infty$ for a fixed $d_\Sigma$ and $K$, then ${\sf vol}({\cal B}_\mathbb{C})\to0$.
\end{proposition}
\begin{IEEEproof}
If $K$ and $d_\Sigma$ are fixed, then the radius of ${\cal B}{_\mathbb R}$ is fixed. From \cite[Lemma 2.5]{John_Info_Age}, ${\sf vol}({\cal B}_\mathbb{R})=\frac{{{\pi ^{{m \mathord{\left/ {\vphantom {m 2}} \right. \kern-\nulldelimiterspace} 2}}}}}{{\Gamma \left( {0.5m + 1} \right)}}{r^m}$ where $m=c_\Sigma d_\Sigma K$. If $c_\Sigma\to\infty$, then ${\sf vol}({\cal B}_\mathbb{R})\to0$. By Bishop-Gromov inequality, ${\sf vol}( {{\cal{B}_\mathbb{C}}})\leq{\sf vol}({{\cal{B}_\mathbb{R}}})$ holds, if ${\cal B}_\mathbb{R}$ and ${\cal B}_\mathbb{C}$ have the same dimension and radius, and a defined Riemannian manifold for ${\cal B}_\mathbb{C}$ has positive Ricci curvature, which is easily guaranteed by following the manifold definition in \cite{Henkel:SpherePacking}. Then ${\sf vol}({\cal B}_\mathbb{C})\to0$ holds by ${\sf vol}({\cal B}_\mathbb{R})\to0$.
\end{IEEEproof}

Thus, if $c_\Sigma$ goes to infinity for a fixed $d_\Sigma$ and $K$, the feasible space of $\mathbf{t}\in{\cal B}_\mathbb{C}$ in (\ref{Norm_Ball}) eventually vanishes.

\section{Numerical Results}

\begin{figure}[!t]
  \centering
   \includegraphics[width=2.80 in]{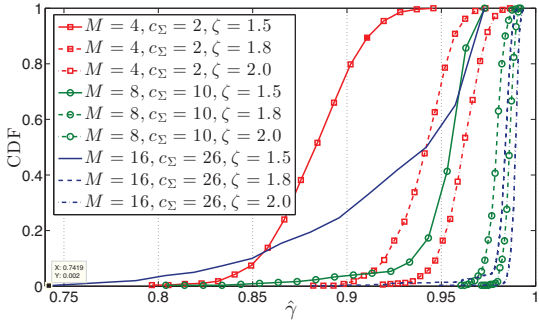}
   \caption{CDF of $\hat\gamma$ for $K=128$, $M=4,8,16$, $N=2$ and $c_\Sigma=c_1+c_2$ where $c_1$ and $c_2$ is the possible maximum value in \eqref{c_j}. }
   \label{Fig_Corollary_R1}
 \end{figure}
\begin{figure}[!t]
  \centering
   \includegraphics[width=3.45 in]{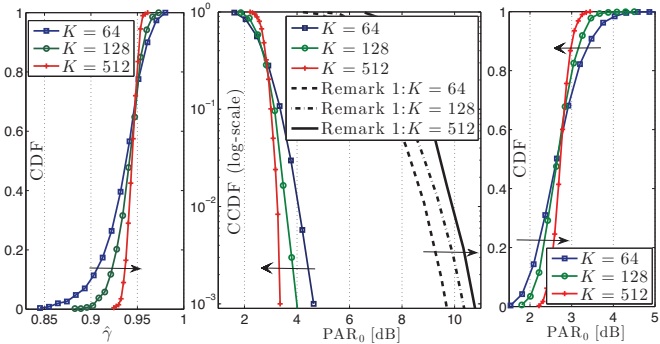}
   \caption{CDF of $\hat\gamma$, CCDF of $\text{PAR}^{(i)}$ and CDF of $\text{PAR}^{(i)}$ (from left to right), where $M=4,c_\Sigma=2$ and $\zeta=1.80$ $(2.55\, \text{dB})$. The arrow direction indicates the increase of $K$.  }
   \label{Fig_Corollary_R2}
\end{figure}

\begin{table}[!t]
\renewcommand{\arraystretch}{1.1}
\caption{Empirical mean of $\text{PAR}^{(i)}$ samples}
\label{Table_Numerical_Result}
\centering
\begin{tabular}{|c|c|c|c|}
\hline   
&    $\zeta=1.5\,$ & $\zeta=1.8\, (2.55 \,\,\text{dB})$  & $\zeta=2.0\,$ \cr \hline
$M=4$ & $1.77$ & $1.90\, (2.79\,\,\text{dB})\,$ & $2.07$  \cr \hline
$M=8$ & $1.54$ & $1.81$ & $2.01$ \cr \hline
$M=16$ & $1.55$ & $1.81$ & $2.01$  \cr \hline     
\end{tabular}
\end{table}

In this section, we numerically evaluate the performance of the proposed scheme assuming $N=J=2$ and $d_1=d_2=1$, and $M$ and $\zeta$ are given in Table \ref{Table_Numerical_Result}, and the channel coefficients and the data streams are respectively drawn from i.i.d. ${\cal CN}(0,1)$ and 16-QAM constellation set $\left(|{\cal A}|=16\right)$. Through large enough realizations of them, we aggregate $\hat{\mathbf{t}}$ and $\gamma$ from $({\cal C})$. The CCDF is defined as ${\sf{Pr}}[\text{PAR}^{(i)}\ge {\text{PAR}}_0]$.

Fig. \ref{Fig_Corollary_R1} plots CDFs of $\hat\gamma$ with respect to various $M$ and $\zeta$, and their corresponding empirical mean of $\text{PAR}^{(i)}$ samples is represented in Table \ref{Table_Numerical_Result}, which shows the effectiveness of the proposed scheme in terms of the average PAR performance. The CDFs of $\hat\gamma$ for $\zeta=1.5$ show that the more redundant spatial dimensions do not always guarantee the better performance, which can be inferred from Proposition \ref{Massive_MIMO}. Thus, $d_\Sigma$ or $K$ needs to be increased with an increase of $M$ to avoid the substantial reduction of the feasible space, especially for quite a challenging constraint such as $\zeta=1.5$.


Fig. \ref{Fig_Corollary_R2} plots the CDFs of $\hat\gamma$, CCDFs of $\text{PAR}^{(i)}$ and CDFs of $\text{PAR}^{(i)}$ from left to right with respect to various $K$ when $M=4$ and $\zeta=1.8$. In this figure, the variance of both $\text{PAR}^{(i)}$ and $\hat\gamma$ decreases as $K$ increases, so that the samples of $\text{PAR}^{(i)}$ and $\hat\gamma$ are increasingly concentrated near their empirical mean value. Thus, the proposed scheme can asymptotically attain $10\log_{10}(\zeta)$ [dB] at $\text{CCDF}=10^{-3}$ as $K$ increases. This result can be inferred from the first discussion item in Section \ref{subsec:discussion}. In Fig. \ref{Fig_Corollary_R2}, the $\text{PAR}^{(i)}$ samples are concentrated near 2.79 [dB] as $K$ grows, thus the proposed scheme can asymptotically attain $\text{PAR}_0=2.79$ [dB] at $\text{CCDF}=10^{-3}$ as $K$ grows. These imply that the variance of $\hat\gamma$ and $\text{PAR}^{(i)}$ when $M=16$ and $\zeta=1.5$ would be reduced if $K$ increases.

The bit error rate (BER) performance degradation compared to the original BD described in Remark \ref{Block_diagonalization} $(\gamma=1)$ can be readily explained, since the received SNR difference of two schemes is expressed as $10\log_{10}(\hat\gamma)$ [dB]. It means that a BER performance curve of the proposed scheme is not shifted more than $10\log_{10}(\hat\gamma)$ [dB] compared to that of Remark \ref{Block_diagonalization}. If $K=64,M=4$ and $\zeta=1.8$, then $\hat\gamma\ge0.8$ as plotted in Fig. \ref{Fig_Corollary_R2}, thus the BER performance degradation is not greater than $10\log_{10}(0.8)\approx0.97$ [dB]. Also, this 0.97 [dB] gets further reduced as $K$ increases, since the minimum value of the $\hat\gamma$ samples increases up to their empirical mean value.


When $\hat\gamma$ is not close to 1, the approximation of $\sqrt{\hat\gamma}$ in our analysis would be inaccurate and lead to larger degradation of the BER performance, so that $\hat\gamma$ of around 1 is of our interest.

\section{Concluding Remarks}
We proposed a MIMO precoding scheme that can achieve the desired PAR. Numerical results show that it can achieve not only the desired PAR on average, but also attains the desired PAR at the $\text{CCDF}=10^{-3}$ if $K$ is arbitrarily large. Also, for the application in a large-scale MIMO system, $\zeta$ and $d_\Sigma$ need to be carefully chosen considering the feasible space size.

\bibliographystyle{IEEEtran}


\bibliography{IEEEabrv,ref_data}

\begin{thebibliography}{10}
\providecommand{\url}[1]{#1}
\csname url@samestyle\endcsname
\providecommand{\newblock}{\relax}
\providecommand{\bibinfo}[2]{#2}
\providecommand{\BIBentrySTDinterwordspacing}{\spaceskip=0pt\relax}
\providecommand{\BIBentryALTinterwordstretchfactor}{4}
\providecommand{\BIBentryALTinterwordspacing}{\spaceskip=\fontdimen2\font plus
\BIBentryALTinterwordstretchfactor\fontdimen3\font minus
  \fontdimen4\font\relax}
\providecommand{\BIBforeignlanguage}[2]{{%
\expandafter\ifx\csname l@#1\endcsname\relax
\typeout{** WARNING: IEEEtran.bst: No hyphenation pattern has been}%
\typeout{** loaded for the language `#1'. Using the pattern for}%
\typeout{** the default language instead.}%
\else
\language=\csname l@#1\endcsname
\fi
#2}}
\providecommand{\BIBdecl}{\relax}
\BIBdecl

\bibitem{Michel:Concentration}
M.~Ledoux, \emph{The concentration of measure phenomenon}.\hskip 1em plus 0.5em
  minus 0.4em\relax Providence, R.I. : American Mathematical Society, 2001.

\bibitem{Donoho:HighDimension}
D.~L. Donoho, ``High-dimensional data analysis: {T}he curses and blessings of
  dimensionality,'' in \emph{Proc. AMS Conference on Math Challenges of the
  21st Century}, 2000.

\bibitem{Larsson:M_MIMO_Megazine}
E.~Larsson, O.~Edfors, F.~Tufvesson, and T.~Marzetta, ``{Massive MIMO for next
  generation wireless systems},'' \emph{{IEEE} Commun. Mag.}, vol.~52, no.~2,
  pp. 186--195, Feb. 2014.

\bibitem{PAPR_Intensified_By_MIMO}
H.-J. Su, C.-P. Lee, W.-S. Liao, R.-J. Chen, C.-L. Ho, C.-L. Tsai, and
  Z.~Yan-Xiu, ``Peak-to-average power ratio issue of beamforming/precoding
  schemes,'' in \emph{IEEE C802.16m-08/1302}, Oct. 2008.

\bibitem{Aggarwal:jd}
A.~Aggarwal and T.~H. Meng, ``{Minimizing the peak-to-average power ratio of
  OFDM signals using convex optimization},'' \emph{{IEEE} Trans. Signal
  Process.}, vol.~54, no.~8, pp. 3099--3110, Aug. 2006.

\bibitem{Studer:en}
C.~Studer and E.~G. Larsson, ``{PAR-aware large-scale multi-user MIMO-OFDM
  downlink},'' \emph{{IEEE} J. Sel. Areas Commun.}, vol.~31, no.~2, pp.
  303--313, Feb. 2013.

\bibitem{Cha:eb}
H.-S. Cha, H.~Chae, K.~Kim, J.~Jang, J.~Yang, and D.~K. Kim, ``{Generalized
  inverse aided PAPR-aware linear precoder design for MIMO-OFDM system},''
  \emph{{IEEE} Commun. Lett.}, vol.~18, no.~8, pp. 1363--1366, Aug. 2014.

\bibitem{BD:Heath_Jr}
R.~Chen, Z.~Shen, J.~Andrews, and R.~Heath, ``Multimode transmission for
  multiuser {MIMO} systems with block diagonalization,'' \emph{{IEEE} Trans.
  Signal Process.}, vol.~56, no.~7, pp. 3294--3302, Jul 2008.

\bibitem{CB_Chae:BD_VecPerturbation}
C.-B. Chae, S.~Shim, and R.~Heath, ``{Block diagonalized vector perturbation
  for multiuser MIMO systems},'' \emph{{IEEE} Trans. Wireless Commun.}, vol.~7,
  no.~11, pp. 4051--4057, Nov. 2008.

\bibitem{John_Info_Age}
J.~Hopcroft and R.~Kannan, ``Foundations of data science,'' \textit{Available
  online only:
  http://research.microsoft.com/en-us/people/kannan/book-dec-30-2013.pdf}, Dec.
  2013.

\bibitem{Barg:Sphere_Exact}
A.~Barg and D.~Nogin, ``Bounds on packings of spheres in the {G}rassmann
  manifold,'' \emph{{IEEE} Trans. Inf. Theory}, vol.~48, no.~9, pp. 2450--2454,
  Sep. 2002.

\bibitem{tao2012_topics}
T.~Tao, \emph{Topics in random matrix theory}.\hskip 1em plus 0.5em minus
  0.4em\relax American Mathematical Society Providence, RI, 2012.

\bibitem{dryden:2005}
I.~L. Dryden, ``Statistical analysis on high-dimensional spheres and shape
  spaces,'' \emph{Ann. Statist.}, vol.~33, no.~4, pp. 1643--1665, Aug. 2005.

\bibitem{Compex_Conv:Heinz}
H.~H. Bauschke and P.~L. Combettes, \emph{Convex Analysis and Monotone Operator
  Theory in Hilbert Spaces}.\hskip 1em plus 0.5em minus 0.4em\relax Springer,
  2011.

\bibitem{Boyd:st}
S.~Boyd and L.~Vandenberghe, \emph{Convex Optimization}.\hskip 1em plus 0.5em
  minus 0.4em\relax Cambridge University Press, 2004.

\bibitem{Henkel:SpherePacking}
O.~Henkel, ``Sphere-packing bounds in the {G}rassmann and {S}tiefel
  manifolds,'' \emph{{IEEE} Trans. Inf. Theory}, vol.~51, no.~10, pp.
  3445--3456, Oct. 2005.

\end{thebibliography}

\end{document}